\documentclass[11pt]{diazessay} 
\usepackage[hidelinks,pdftex, pdfkeywords={paglen, crawford, excavating ai, prada, training humans, informed consent, digital ethics}]{hyperref}
\hypersetup{
  colorlinks   = true, 
  urlcolor     = blue, 
  linkcolor    = blue, 
  citecolor   = red 
}
\newcommand{\q}[1]{``#1''}

\title{\textbf{Excavating \q{Excavating AI}: \\{The Elephant in the Gallery}}}

\author{\textbf{Michael J. Lyons} \\ \textit{Ritsumeikan University}} 

\date{}

\begin{document}

\maketitle 



\begin{abstract}
\noindent Two art exhibitions, \q{Training Humans} and \q{Making Faces,} and the accompanying essay \q{Excavating {AI}: The politics of images in machine learning training sets} by Kate Crawford and Trevor Paglen, are making substantial impact on discourse taking place in the social and mass media networks, and some scholarly circles. Critical scrutiny reveals, however, a self-contradictory stance regarding informed consent for the use of facial images, as well as serious flaws in their critique of ML training sets. Our analysis underlines the non-negotiability of informed consent when using human data in artistic and other contexts, and clarifies issues relating to the description of ML training sets.
\end{abstract}
\hspace*{3.6mm}\textit{Keywords:} digital ethics, informed consent, training sets, ai, \\ \hspace*{20mm} machine learning, affective computing  

\vspace{30pt} 



\noindent
In this essay, I will present a case study and critical analysis involving two art exhibitions and an accompanying essay by contemporary artist Trevor Paglen and media studies scholar Kate Crawford. \q{\href{http://www.fondazioneprada.org/project/training-humans/?lang=en}{Training Humans},} held at the Milan Osservatorio, Fondazione Prada, from September 2019 to February 2020, exhibited collections of images of human faces used for training computer vision systems. An associated exhibition, \q{\href{https://t.co/T4wXYStZGS}{Making Faces},} was organized by Prada Mode Paris in January 2020 to coincide with the opening of the Paris Haute Couture event. An essay by Crawford and Paglen, \q{Excavating {AI}: The politics of images in machine learning training sets,} was self-published online to accompany the exhibitions \cite{crawford2019excavating}.\footnote{Henceforth, the abbreviations C\&P, TH, MF, EAI will refer respectively to the authors, the two exhibitions, and the essay.}

According to the Fondazione Prada \href{http://www.fondazioneprada.org/project/training-humans/?lang=en}{web page}, the exhibition explored:
\begin{quote}
\textit{
\ldots two fundamental issues in particular: how humans are represented, interpreted and codified through training datasets, and how technological systems harvest, label and use this material. 
}
\end{quote}
Though sympathetic to C\&P's overall aims, several issues troubled me about the exhibition, TH, its satellite, MF, and the analysis in EAI \cite{crawford2019excavating}. Briefly, the main concerns are:
\begin{itemize}
    \item C\&P's description of machine learning training sets is inadequate and flawed, containing both generic and specific errors
	\item C\&P employed ethical double-standards by exhibiting images of private individuals without informed consent.
	\item C\&P failed to respect clearly stated terms of use for several of the datasets.
	\item The EAI essay contains important factual errors and misleading statements
\end{itemize}
The present commentary explains and elaborates on these concerns to propose that C\&P's flawed approach compromises their aims. This critical analysis and commentary is intended to  contribute constructively to dialogue concerning the use of human data for artistic and other purposes and to correct some of the mistakes that have been instigated by the exhibitions and the essay, EAI.
\begin{figure}[t]
\begin{center}
	\includegraphics[width=.8\linewidth]{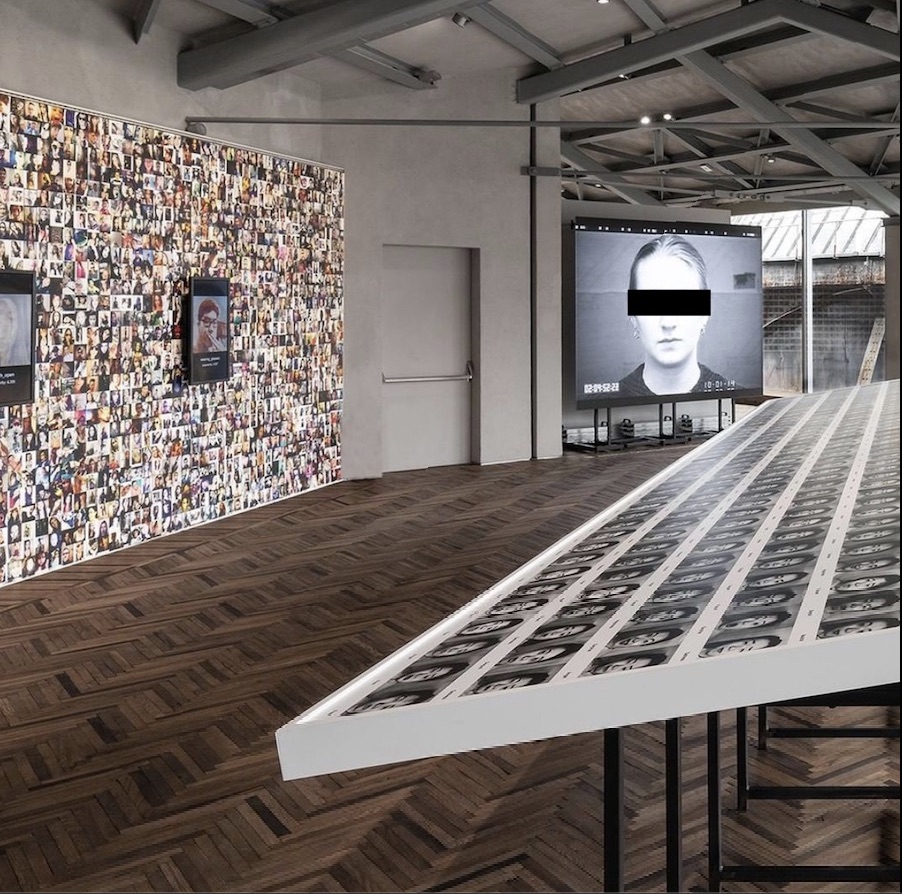}
	\caption{\q{Training Humans} exhibition, Milan Osservatorio.}
\label{thprada}
\end{center}
\end{figure}

\section*{Disclosure}
The following factors have informed and influenced my comments:
\begin{itemize}
	\item I am a co-author, with colleagues Miyuki Kamachi and Jiro Gyoba, of the JAFFE image set,\footnote{The \href{https://zenodo.org/record/3451524}{JApanese Female Facial Expressions}, a set of images visible on the right in Figure \ref{thprada}.} that was exhibited at TH, MF, and discussed in EAI.
	\item I worked briefly on face recognition technology, as manager of a team that took part in Phase II of the FERET face recognition competition held in March 1995. I was not comfortable working on surveillance technology and soon left the field. Since that time, for various reasons, I have been opposed to the deployment of surveillance systems.
	\item I have a long term interest in contemporary art and have encountered some of Trevor Paglen's projects at exhibitions and in books. My opinion of his past work has been relatively positive.
	\item There are no competing financial interests.
\end{itemize}

\section*{Training Sets and False Taxonomy}
Crawford and Paglen's two exhibitions, TH and MF, and the essay EAI, may be viewed as a critique of image taxonomy, and especially as a caveat concerning the political implications of labelling photographs of human individuals. Most spectacularly, their project popularized bizarre and demeaning labels on some of the person categories in the ImageNet database. 

The foundation of C\&P's analysis of computer vision training sets, however, is itself marred by a taxonomic blunder. The fundamental problem is that C\&P attempt to subsume a very heterogeneous selection of datasets into the single undifferentiated category of machine learning 'training set.' The datasets exhibited at C\&P are diverse with regard to origin, intended purpose, status of copyright and informed consent, terms of use, funding source etc.

In the following I will illustrate the importance of making distinctions between various kinds of image datasets.
C\&P exhibited facial image datasets originating in two distinct ways: sets that were carefully designed and photographed by research groups under controlled laboratory conditions, and sets consisting of images scraped in bulk from the internet. I refer to these respectively as \textit{constructed datasets}  and \textit{scraped datasets}. 

Consider how their different origin affects the public exhibition of the images. There are, of course, ethical concerns with the unauthorized public exhibition of both kinds of image datasets, but with an important distinction: the status of copyright and informed consent are precisely known for the constructed datasets, and uncertain or unknown for the scraped datasets.

In contrast with scraped training sets, constructed image sets such as JAFFE, FERET,\footnote{\href{https://www.nist.gov/programs-projects/face-recognition-technology-feret}{FacE REcognition Technology dataset \cite{phillips2000feret}}} and CK\footnote{\href{http://www.jeffcohn.net/Resources/}{Cohn-Kanade facial expression dataset \cite{kanade2000comprehensive}}, visible at the back in Figure \ref{thprada}.} also have explicitly defined \textit{terms of use}. These three sets permit use for the purpose of \textit{non-commercial scientific research}, and allow for limited reproduction of the images in scholarly articles reporting research results.

By exhibiting the images publicly in art shows, C\&P breached the terms of use for the constructed sets JAFFE, CK, and FERET. The artists and the Fondazione Prada claim that their use did constitute `non-commercial scientific research'\footnote{Letter from the Fondazione Prada, Aug 6, 2020.}, but this entails an untenable semantic stretch. 

Consider, for example, the description for the satellite exhibition \q{Making Faces,} held as part of Prada Mode
\begin{quote}
\textit{
\ldots a travelling social club with a focus on contemporary culture that provides members a unique cultural experience along with music, dining, and conversations \ldots
}
\end{quote}
held at the exclusive Belle Époque restaurant Maxim's featuring `exhilarating performances,' celebrity guests, and a two-day `food experience' curated by a Michelin-starred chef. I can only guess the budget for this luxury junket, but do know that the image sets defining the themed event cost them not a single euro cent. 
\begin{figure}[t]
\begin{center}
	\includegraphics[width=.8\linewidth]{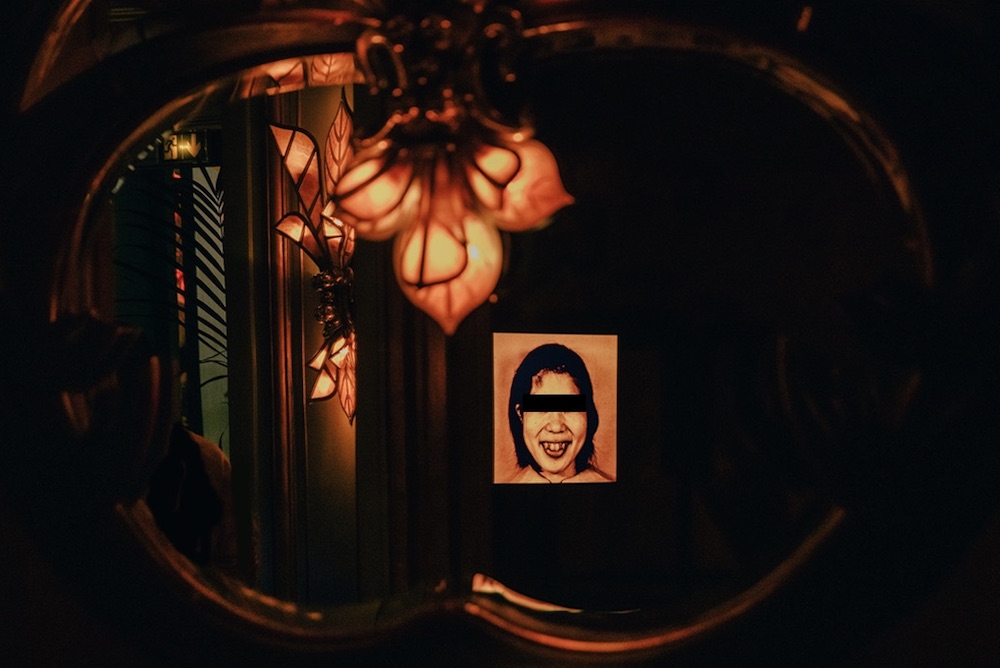}
	\caption{\q{Making Faces} exhibition at Maxim's, Prada Mode Paris.}
\label{default}
\end{center}
\end{figure}

The association of the Prada fashion brand with TH and MF further calls to question C\&P's claim of non-commercial status. The Fondazione Prada, which hosted TH, describes itself as a `non-profit organization with a mission to promote and encourage cultural fields such as contemporary art.' However, holding \q{Making Faces} as part of the  ostentatious Prada Mode Paris fashion event, with the visible participation of the Prada Group CEO, points to a tighter association with corporate interests. The close participation of a private corporation in an exhibition intended to probe what is essentially a matter of public policy, \textit{viz.} the politics of surveillance technology, should itself beg scrutiny. What vested interests might such a corporate sponsor hold? Can such activities be sincerely described as devoid of commercial interest?

More concretely, entry to TH required purchasing a ticket, a commercial transaction. There is an exhibition catalogue (containing JAFFE images) for sale\footnote{\href{http://bookshop.fondazioneprada.org/shop/UIShop/Shop_DettaglioArticolo.aspx?idshoparticolo=117}{Quaderni Fondazione Prada \#26: Training Humans, ISSN 2421-2792}} via the museum bookshop. It is not unreasonable to suppose that the exhibitions may have boosted sales of Paglen's photobooks or otherwise helped to move merchandise. 

Financial transactions aside, TH/MF created a media spectacle \cite{tedone2019from}, attracting considerable public attention, often considered a boon by artists and academics alike. The bottom line is that one cannot honestly describe TF or MF as `scientific research' and both took place in contexts that were far from non-commercial.

There is an uncanny resemblance to processes described by Shoshana Zuboff  in her analysis of \textit{surveillance capitalism} \cite{zuboff2019age}. Datasets mined online, a \q{surplus resource} obtained at no cost, were repackaged as commodities for consumption by the contemporary art world and popular media, yielding considerable rewards, financial and otherwise. C\&P might argue that their work is intended to render the public a beneficial service. Whether or not such a rationale is justifiable, it strangely echos the alibis of the behemoth surveillance capitalists. 

\section*{Informed Consent and Digital Ethics Malpractice}
Scientific researchers who conduct experiments involving humans are required to provide their subjects with knowledge sufficient to make an informed, voluntary decision to participate or not \cite{10.2307/29733727}. This principle, known as \textit{informed consent}, also concerns any personal data, including photographs, acquired in research: subjects must be informed about how their personal data will be used and disseminated, and voluntarily provide agreement \cite{nature2019time}.

All three constructed datasets were assembled in projects regulated by the human subjects ethics committees of their respective institutions, requiring the informed consent of the private individuals shown in the images and videos. All three sets prohibit redistribution. The FERET and CK pages ask potential users to download, sign, and submit an application form before receiving access to the images. JAFFE, until recently, relied on an honour code, asking users to read and agree to conditions of use before downloading.\footnote{Until now, there were no serious abuses of this policy in more than twenty years. Access is now restricted and vetted.} These policies are intended, in part, to protect the privacy of the depicted individuals, according to the conditions under which they gave informed consent.

Whatever one may think about C\&P's mishandling of terms of use for the constructed image datasets in TH/MF, shirking the requirement for informed consent is a more serious matter. The authors of JAFFE, CK, and FERET obtained informed consent contingent on the terms of use. When they downloaded the datasets, C\&P certainly did not acquire the right to modify those terms to suit their ends. To exhibit these images in public, C\&P should have first obtained permission from the persons depicted. In an interview with Gaia Tedone,\footnote{\href{https://youtu.be/BSaYROvz29E}{\q{What are the ethics of exhibiting datasets of faces?}} Interview by Gaia Tedone \cite{tedone2019from}.} Paglen seems to be aware of ethical complications associated with exhibiting photographs of private persons without their knowledge, and Crawford adds: 
\begin{quote}
\textit{
\ldots we were really careful to create the entire exhibition as a thing where if you see a face that you know or yourself that you can actually choose to have it removed and this is a freedom and a sense of agency that doesn't normally exist \ldots
}
\end{quote}
This strongly resembles the promise IBM makes for it's non-consensual \q{Diversity in Faces} dataset \cite{solon2019facial}. Unfortunately, informed consent makes no sense \textit{ex post facto}---to have any meaning at all informed consent must be obtained \textit{beforehand}. 

There are similarities with two types of digital ethics malpractice identified by Luciano Floridi \cite{floridi2019translating}. Slyly twisting the principle of informed consent to fit one's convenience is an instance of \textit{ethics shopping}. Offering to redress the appropriation of personal images, after the fact, is an instance of \textit{ethics bluewashing}. To posit, however, that what is undeniably non-consensual use endows the victim with exceptional `freedom and agency,' is a mind-bending  masterpiece of sophistry.\footnote{\href{https://youtu.be/BSaYROvz29E}{Watch the video} to experience the full impact.}
\begin{figure}[t]
\begin{center}
	\includegraphics[width=.7\linewidth]{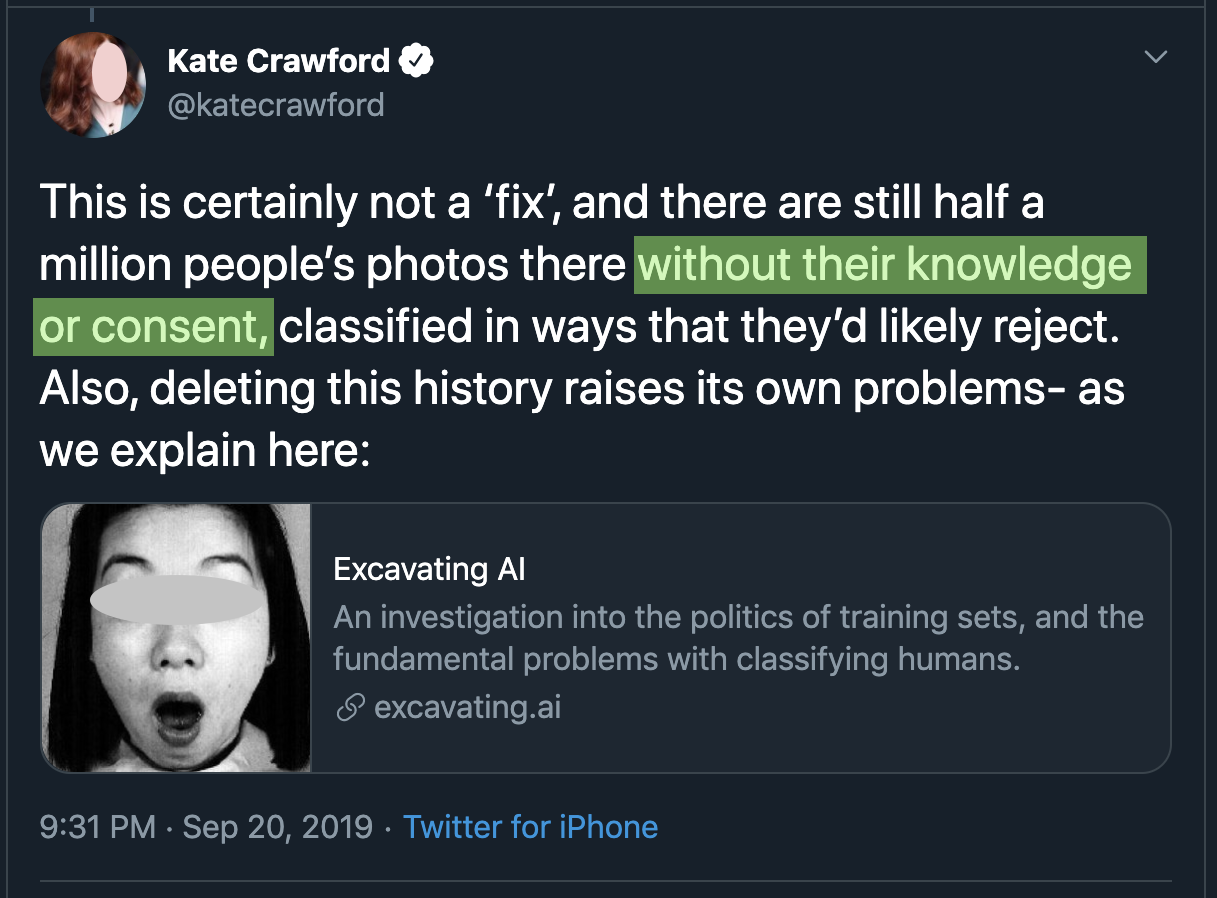}
	\caption{An astonishing tweet.}
\label{crawfordtweet}
\end{center}
\end{figure}
Scraped datasets contain images of varying copyright status, but researchers navigate this barrier by invoking the principle of \textit{fair use} or \textit{fair dealing}. For photographs and videos of private individuals, however, fair use and even the permissive terms of a creative commons license, refer only to copyright and do not satisfy the requirement for informed consent \cite{solon2019facial, prabhu2020large}. In EAI, C\&P underline the issue, when they comment on \begin{quote}
\textit{
\ldots the practice of collecting hundreds of thousands of images of unsuspecting people who had uploaded pictures to sites like Flickr \ldots
}
\end{quote}

Kate Crawford has tweeted about the lack of informed consent for the use of private images in scraped training sets (see Figure \ref{crawfordtweet}). Remarkably, however, Crawford does not seem to have noticed that she did not have the informed consent of the woman whose face she was tweeting. The irony is acute---this JAFFE image was used, without permission or informed consent, as the icon for the EAI web page: it was embedded automatically with every post of the EAI site to social media. In effect, the EAI site acted as a machine that caused anyone who shared it on social media to unwittingly breach informed consent and the JAFFE terms of use.\footnote{As of Aug 30, 2020 the site icon had changed, possibly in response to my request. The screenshot in figure \ref{crawfordtweet} is from Aug 23, so the image seems to have been used this way for more than 11 months. I have masked the faces.} 
\begin{figure}[t]
\begin{center}
	\includegraphics[width=.9\linewidth]{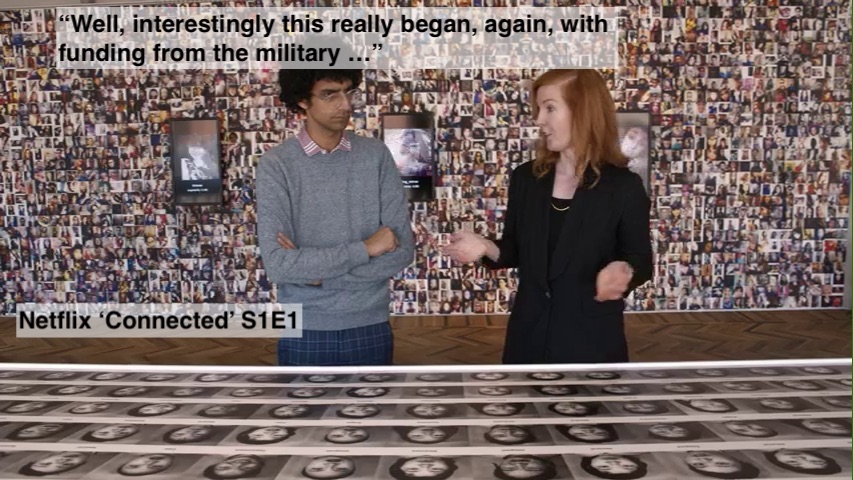}
	\caption{An unsupported claim.}
\label{netcon}
\end{center}
\end{figure}
\section*{Social Media Fallout}
Am I taking issue with a mere technicality? For the JAFFE images, at least, the non-consensual public exhibition had several unwelcome consequences. Soon after TH opened in Sept. 2019, photos of the JAFFE images, by professional photographers and museum visitors with smartphones, began to multiply.\footnote{Judging by the dates of the posts---I did not see these until many months later.} Photos showed up on social network sites like Instagram, Twitter, and Facebook; in news reports and online magazines; in videos on YouTube and Vimeo; on the Fondazione Prada website; on the EAI web page, and in many other places. The proliferation of such photos escalated after MF. A photograph clearly showing the face of one JAFFE woman, now a successful professional with a public persona, appeared for sale at Getty Images with a price tag attached. When I alerted that woman to the situation, she was shocked and dismayed. The JAFFE volunteers certainly did not consent to such indiscriminate dissemination of their photographs. 

I noticed the widespread proliferation of JAFFE images in August 2020, long after TH/MF had closed. Too late for that golden promise of exceptional `freedom and agency.' My attempts to undo the damage caused by the TH/MF media spectacle proved time-consuming, frustrating, and ultimately impossible. I was able, at least, to convince Getty Images to take down the offending photograph, but only after a multi-day effort that needed several emails and intercontinental phone calls. 

Views of the JAFFE and CK faces and were broadcast as part of a Netflix science-lite documentary 'Connected' in an episode titled 'Surveillance.' The production company, Zero Point Zero, had contacted me to ask for, and did not receive, my permission to show the images. They did not heed the negative response, nor has Zero Point Zero admitted that this was a serious breach of journalistic ethics. To add insult to the injury, while discussing the images with show host Latif Nasser, Kate Crawford fabulates a narrative attributing military origins and funding to the research (see Figure \ref{netcon}). Months have passed, but Crawford has still not responded to my request to provide an explanation for this groundless claim.

\section*{Ceci n'était pas un `Training Set'}
The blurb at the  TH exhibition web page states: \begin{quotation}\q{Training Humans} explores two fundamental issues in particular: how humans are represented, interpreted and codified through training datasets, and how technological systems harvest, label and use this material.\footnote{\href{http://www.fondazioneprada.org/project/training-humans/?lang=en}{Training Humans}, Milan Osservatorio, Fondazione Prada.}\end{quotation}

We have seen that JAFFE and some other image sets were not `harvested' by a `system' but carefully designed and constructed by researchers. The quote implies another fallacy: that JAFFE is primarily a `training dataset.' C\&P's claim that JAFFE was intended to be a `training dataset' is pure fiction. As it turns out,  JAFFE began with the scientific aim of modelling data on facial expression perception by humans. 

C\&P say that TH/MF took two years of research to prepare. It appears, however, that they did not find the time to read the documents attached to JAFFE: a \q{README\_FIRST.txt} file, that contains an explicit caveat regarding the image labels, and an article describing how the image set was assembled and how it was used \cite{lyons1998coding}. The article's title alone, \q{Coding Facial Expressions with Gabor Wavelets,} does not mention recognition, classification, or learning. That should have acted as a clue. Had they looked at the article, C\&P might have been puzzled to see that no machine learning algorithms were trained, no images classified, and no recognition rates reported.

Let's look more closely at C\&P's mistaken account of JAFFE:
\begin{quotation}
\textit{The intended purpose of the dataset is to help machine-learning systems recognize and label these emotions for newly captured, unlabelled images. The implicit, top-level taxonomy here is something like `facial expressions depicting the emotions of Japanese women.'}
\end{quotation}
C\&P wrongly project onto JAFFE a purpose that was not what we had in mind when we designed and photographed the image set in 1996. Likewise, we have never claimed, anywhere, that the photos represent felt emotions---we describe the images unambiguously as posed facial expressions. EAI continues:
\begin{quotation}
\textit{\ldots there’s a string of additional assumptions \ldots that there are six emotions plus a neutral state; that there is a fixed relationship between a person’s facial expression and her true emotional state; and that this relationship between the face and the emotion is consistent, measurable, and uniform across the women in the photographs.}
\end{quotation}
Neither have we nor has Ekman, ever claimed that there are only six emotions. This is an elementary misconception about Ekman's work that a basic familiarity with the introductory facial expression literature could have helped C\&P avoid, here, and in several spoken presentations. Lisa Feldman Barrett, a prominent Ekman critic cited by C\&P, has conducted experiments using five facial expressions plus a neutral face \cite{gendron2014perceptions}. Should we conclude, using C\&P's reasoning, that Barrett believes there are only five emotions? 

From, for example, the critical historical account given by Ruth Leys \cite{leys2017ascent}, which EAI cites, C\&P could have learned that Ekman's proposal refers to six universally recognized \textit{basic facial expressions} (BFE) plus any number of culturally variable facial expressions \cite{ekman1992argument}. Ekman has written elsewhere that there may be thousands of facial expressions, in addition to his basic six \cite{ekman2003emotions}. C\&P list several further assumptions that we supposedly made about felt emotion. Unfortunately, these are also made without any justification.

\section*{Why we Made JAFFE}
What, then, was the intended use of the JAFFE dataset? The project had two non-technological, scientific aims: 
\begin{itemize}
	\item to test the psychological plausibility of a biologically inspired model \cite{lyons1996model, lyons1997v1,lyons1997gabor,lyons2000linked} for facial expression representation.
	\item to explore the relationship between categorical and dimensional paradigms in facial expression research \cite{lyons1997gabor2, lyons1998coding, lyons2007dimensional}. 
\end{itemize}	
Very briefly, emotions have been characterized as \textit{categorical} by Darwin \cite{darwin1872express}, Tomkins \cite{tomkins1962affect}, and Ekman \cite{ekman1993facial}, and as \textit{dimensional} by Wundt \cite{wundt1897outline}, Schlosburg \cite{schlosberg1952description}, Russell \cite{russell1980circumplex}, and Barrett \cite{barrett2006emotions}. Using our \q{Linked Aggregate Code} \cite{lyons2000linked} and non-metric multidimensional scaling (nMDS), a statistical technique, we discovered evidence for low dimensional structure derived from visual aspects of the BFE images. Without referring to labels or semantic evaluations, we recovered arrangements of the BFE resembling the \q{affective circumplex.} Effectively this suggested a way to bridge the competing categorical and dimensional models whose roots both date to the beginnings of scientific psychology (Wundt and Darwin). The discovery was confirmed a few years later by Dailey and Cottrell, using a different, but related, approach \cite{dailey2002empath}.

\section*{How JAFFE Became a `Training Set'}
In the interests of an open data policy, we began to provide the JAFFE images and semantic ratings to other scientific researchers.\footnote{C\&P did not acknowledge that both JAFFE and CK datasets have been widely used in experimental and computational psychology, neuroscience, and other areas not related to machine learning.} After attending one of my talks, Zhengyou Zhang, a computer vision researcher, initiated a collaboration on automatic facial expression classification \cite{zhang1998comparison}. That was the first use of JAFFE as a `training set.' Subsequently, many pattern recognition studies have used JAFFE as a benchmark for comparing classification algorithms. JAFFE may have \textit{become} a `training set,' but we did not create it with this intention, as C\&P claim in EAI. 

Most such studies use JAFFE not only for training but also for validating and testing algorithms. C\&P's presentation naïvely implies that an algorithm is trained using a dataset like JAFFE, then it is unleashed on the real world.  Indeed the machine learning studies using JAFFE are typically academic `toy world' studies that are not deployable in real-world applications.

C\&P continue:
\begin{quote}
\textit{
The JAFFE training set is relatively modest as far as contemporary training sets go. It was created before the advent of social media, before developers were able to scrape images from the internet at scale.
}
\end{quote}
Their narrative (plainly wrong, again)  is that we would have scraped the social networks for facial images had these been available. However, photos with uncontrolled lighting, pose, camera, of subjects, wearing makeup, and jewelry, would not have served our scientific purposes. On the other hand, it would not have been difficult to photograph more volunteers if we had needed to---but the small number of JAFFE posers was sufficient for the original study. Had we set out to build a facial expression `training set,' we would and certainly could  have photographed a larger and more diverse group of people. 

Again, JAFFE was not intended for training machine learning algorithms. By using JAFFE as the `anatomical model' for their exposition of training set `taxonomy,' C\&P based their discourse on a taxonomy of datasets that is itself mistaken.

\section*{Mind-Reading Machines}
I doubt that many of the engineers and computer scientists who use JAFFE for machine learning research have ever thought much about the psychology of human emotion. Most are probably not aware of the vast and complex literature relating to facial expression. I first noticed this in the collaboration with Zhang. During the preparation of the joint publication, Zhang was eager to claim that the results proved the neural network could recognize facial expressions with greater accuracy than humans---a misunderstanding of the semantic ratings experiments. I had to veto this mistaken claim from the co-authored article \cite{zhang1998comparison}. 

Regarding C\&P's fabulation that we aimed to build a machine that reads minds from faces, I have only never discussed `mind-reading machines' except to express my profound skepticism of such projects \cite{lyons2007dimensional}.

\section*{Critique of Ekman's Work}
On the TH web page, we find the following oversimplified and factually wrong description of Paul Ekman's work:
\begin{quotation}
\textit{Based on the heavily criticized theories of psychologist Paul Ekman, who claimed that the breadth of the human feeling could be boiled down to six universal emotions,}
\end{quotation}
C\&P also present facile, and even derisive-sounding accounts of Ekman's work in public talks that are misleading for the unsuspecting.\footnote{For example, \href{https://youtu.be/LPbtYG3iZAk}{this talk}, Haus der Kulturen der Welt, Berlin, Jan 12, 2019} They are not wrong that Ekman's views are contested. That is nothing new: when Ekman entered the field in the 1960s, he encountered opposition from the then dominant social-constructionists Margaret Mead and Ray Birdwhistell \cite{leys2017ascent}. From the 1980s, there has been a prolonged debate involving Ekman and James Russell, Alan Fridlund, and others \cite{russell2003core,fridlund1994human}. 

Despite the impression conveyed by C\&P, however, it is not accurate to describe Ekman's work as discredited. A survey conducted by Ekman in cooperation with his opponent James Russell, and discussed by Alan Fridlund, found acceptance of some of Ekman's main views among a majority of researchers working on the psychology of affect \cite{ekman2016scientists, crivelli2018facial}. Lisa Barrett, now leading the critique of Ekman's work and, more generally, the `standard model,' can be interpreted as recognition of a need for a radical paradigm shift in emotion research \cite{barrett2017emotions}. This important critical work has gained attention and interest, but it cannot be said to have established a new standard paradigm \cite{adolphs2019emotion}.

Whether or not we accept Ekman's views,\footnote{My own views combine dimensional/categorical and nativist/cultural aspects.} we should recognize that throughout his long career, he has tested his ideas experimentally, published findings in peer-reviewed articles, and engaged in vigorous open debates with opponents. With this in mind, C\&P's uninformed and biased\footnote{In the sense that C\&P mention only Ekman's harshest critics, seemingly without fully understanding the criticism, while neglecting other viewpoints \cite{heaven2020faces, cowen2019mapping, elfenbein2002universality, srinivasan2018cross}.}  portrayal of Ekman's views is regrettable. 

\section*{Summary: Errors, Ethics, Constraints, and Creativity}
\begin{quote}
\textit{
In disputes upon moral or scientific points, let your aim be to come at truth, not to conquer your opponent. So you never shall be at a loss in losing the argument, and gaining a new discovery.
\begin{flushright}Arthur Martine, 1866 \cite{martine1866martine}\end{flushright}
}
\end{quote}

C\&P's essay \q{Excavating AI} frames their analysis of `training sets' in terms of a grand archaeological metaphor, to signify their method of 
\begin{quote}
\textit{
\ldots digging through the material layers, cataloging the principles and values by which something was constructed \ldots
}
\end{quote}

In the title of this document, I reuse the verb `to excavate'  more modestly: I intended to dig in and examine how C\&P's analysis of `training sets' holds up to scrutiny. Though my comments are not  exhaustive,\footnote{A full discussion on the `echos of phrenology' trope is beyond the scope of this commentary. Briefly, EAI does not sufficiently acknowledge the existing, well-documented controversy over the revival of physiognomy \cite{highfield2009your,wang2018deep}. Overall, I am skeptical of the attempt to impose a grandiose narrative.}
 I have uncovered faulty analysis, elementary errors, misunderstandings, and questionable reasoning. 
 By choosing JAFFE as the `anatomical model'  for their exposition of training set `taxonomy,' without having made sufficient effort to understand what it is, C\&P rashly compromised the core of their discourse. This key error amounts to faulty taxonomy, appropriately enough. By attacking a distinguished scholar, apparently without having studied his writings,   C\&P raised doubts about their own scholarship. C\&P wrote:
\begin{quote}
\textit{
\ldots when we look at the training images widely used in computer-vision systems, we find a bedrock composed of shaky and skewed assumptions
}
\end{quote}
not realizing that shaky and skewed is a fitting description of the `Excavating AI' essay itself. To be sure, computer vision is a technically challenging field. The literature of facial expression research is complex and confusing for the uninitiated. Perhaps C\&P underestimated the difficulties of crossing disciplines. EAI offers a compelling narrative---for readers not knowledgeable or critical enough to recognize the fallacies. 

I began this commentary with an assertion of sympathy for C\&P's aims. That has not changed. Surveillance technologies must be monitored and regulated via open, democratic policies. Corporations must defer to the primacy of human rights. A good starting point is expressed clearly in article one of the Nuremberg Code \cite{10.2307/29733727}. Informed voluntary consent is essential and non-negotiable when dealing with human data. 

By exhibiting images from constructed datasets such as JAFFE and CK, without first obtaining informed consent, Crawford and Paglen demonstrated what is, for my tastes, an insufficient level of respect for this fundamental human right. The flaws in EAI may be disappointing, but the failure to observe the necessity for informed consent reveals an egregious ethical double standard---the elephant in the gallery at \q{Training Humans} and \q{Making Faces.} If Crawford and Paglen are willing to overlook informed consent, why should they expect anyone to do otherwise?

Is there no way to investigate the aesthetics of facial image training sets without violating informed consent? Creativity is said to thrive in the presence of constraints. Well, here is a constraint related directly to what really is at stake---human agency, freedom, dignity. Surely informed consent is something worth working \textit{with}, not against? 

Kate Crawford and Trevor Paglen have yet to acknowledge the serious errors contained in \q{Excavating AI} or their self-contradictory stance regarding informed consent. Neither have they admitted the negative consequences of their actions. I hope they will eventually realize the importance of doing so.   

\vspace{5 mm}
{\noindent\small \textbf{Michael Lyons is Professor of Image Arts and Science at Ritsumeikan University}}

\bibliographystyle{acm}

\bibliography{eeai3.bib}

\begin{thebibliography}{10}

\bibitem{10.2307/29733727}
The {N}uremberg {C}ode (1947).
\newblock {\em BMJ: British Medical Journal 313}, 7070 (1996), 1448--1448.

\bibitem{adolphs2019emotion}
{\sc Adolphs, R., Mlodinow, L., and Barrett, L.~F.}
\newblock What is an emotion?
\newblock {\em Current Biology 29}, 20 (2019), R1060--R1064.

\bibitem{barrett2006emotions}
{\sc Barrett, L.~F.}
\newblock Are emotions natural kinds?
\newblock {\em Perspectives on psychological science 1}, 1 (2006), 28--58.

\bibitem{barrett2017emotions}
{\sc Barrett, L.~F.}
\newblock {\em How emotions are made: The secret life of the brain}.
\newblock Houghton Mifflin Harcourt, 2017.

\bibitem{cowen2019mapping}
{\sc Cowen, A., Sauter, D., Tracy, J.~L., and Keltner, D.}
\newblock Mapping the passions: Toward a high-dimensional taxonomy of emotional
  experience and expression.
\newblock {\em Psychological Science in the Public Interest 20}, 1 (2019),
  69--90.

\bibitem{crawford2019excavating}
{\sc Crawford, K., and Paglen, T.}
\newblock Excavating {AI}: The politics of images in machine learning training
  sets.
\newblock [Online; accessed 27-August-2020].

\bibitem{crivelli2018facial}
{\sc Crivelli, C., and Fridlund, A.~J.}
\newblock Facial displays are tools for social influence.
\newblock {\em Trends in Cognitive Sciences 22}, 5 (2018), 388--399.

\bibitem{dailey2002empath}
{\sc Dailey, M.~N., Cottrell, G.~W., Padgett, C., and Adolphs, R.}
\newblock Empath: A neural network that categorizes facial expressions.
\newblock {\em Journal of cognitive neuroscience 14}, 8 (2002), 1158--1173.

\bibitem{darwin1872express}
{\sc Darwin, C.}
\newblock {\em The expression of the emotions in man and animals.}
\newblock John Murray, 1872.

\bibitem{nature2019time}
{\sc Editorial}.
\newblock Time to discuss consent in digital-data studies.
\newblock {\em Nature 572}, 7767 (2019), 5.

\bibitem{ekman1992argument}
{\sc Ekman, P.}
\newblock An argument for basic emotions.
\newblock {\em Cognition \& emotion 6}, 3-4 (1992), 169--200.

\bibitem{ekman1993facial}
{\sc Ekman, P.}
\newblock Facial expression and emotion.
\newblock {\em American psychologist 48}, 4 (1993), 384.

\bibitem{ekman2003emotions}
{\sc Ekman, P.}
\newblock {\em Emotions revealed: Recognizing faces and feelings to improve
  communication and emotional life.}
\newblock Times Books, 2003.

\bibitem{ekman2016scientists}
{\sc Ekman, P.}
\newblock What scientists who study emotion agree about.
\newblock {\em Perspectives on psychological science 11}, 1 (2016), 31--34.

\bibitem{elfenbein2002universality}
{\sc Elfenbein, H.~A., and Ambady, N.}
\newblock On the universality and cultural specificity of emotion recognition:
  a meta-analysis.
\newblock {\em Psychological bulletin 128}, 2 (2002), 203.

\bibitem{floridi2019translating}
{\sc Floridi, L.}
\newblock Translating principles into practices of digital ethics: Five risks
  of being unethical.
\newblock {\em Philosophy \& Technology 32}, 2 (2019), 185--193.

\bibitem{fridlund1994human}
{\sc Fridlund, A.~J.}
\newblock {\em Human facial expression: An evolutionary view.}
\newblock Academic Press, 1994.

\bibitem{gendron2014perceptions}
{\sc Gendron, M., Roberson, D., van~der Vyver, J.~M., and Barrett, L.~F.}
\newblock Perceptions of emotion from facial expressions are not culturally
  universal: evidence from a remote culture.
\newblock {\em Emotion 14}, 2 (2014), 251.

\bibitem{heaven2020faces}
{\sc Heaven, D.}
\newblock Why faces don't always tell the truth about feelings.
\newblock {\em Nature 578}, 7796 (2020), 502--504.

\bibitem{highfield2009your}
{\sc Highfield, R., Wiseman, R., and Jenkins, R.}
\newblock How your looks betray your personality.
\newblock {\em New Scientist 201}, 2695 (2009), 28--32.

\bibitem{kanade2000comprehensive}
{\sc Kanade, T., Cohn, J.~F., and Tian, Y.}
\newblock Comprehensive database for facial expression analysis.
\newblock In {\em Proceedings Fourth IEEE International Conference on Automatic
  Face and Gesture Recognition (Cat. No. PR00580)\/} (2000), IEEE, pp.~46--53.

\bibitem{leys2017ascent}
{\sc Leys, R.}
\newblock {\em The ascent of affect: Genealogy and critique}.
\newblock University of Chicago Press, 2017.

\bibitem{lyons1998coding}
{\sc Lyons, M., Akamatsu, S., Kamachi, M., and Gyoba, J.}
\newblock Coding facial expressions with gabor wavelets.
\newblock In {\em Proceedings Third IEEE international conference on automatic
  face and gesture recognition\/} (1998), IEEE, pp.~200--205.

\bibitem{lyons1997gabor2}
{\sc Lyons, M., Kamachi, M., Gyoba, J., and Akamatsu, S.}
\newblock Gabor wavelet representation of facial expression.
\newblock {\em Technical report of IEICE. HIP 97}, 117 (jun 1997), 9--15.

\bibitem{lyons1997v1}
{\sc Lyons, M., Kamachi, M., Tran, P., Gyoba, J., and Akamatsu, S.}
\newblock V1 similarity measure recovers dimensions of facial expression
  perception.
\newblock {\em Investigative Ophthalmology \& Visual Science 38}, 4 (1997).

\bibitem{lyons1996model}
{\sc Lyons, M., and Morikawa, K.}
\newblock A model based on {V}1 cell responses predicts human perception of
  facial similarity.
\newblock {\em Investigative Opthalmolgy and Visual Science 37}, 910 (1996).

\bibitem{lyons1997gabor}
{\sc Lyons, M., Morikawa, K., and Akamatsu, S.}
\newblock Gabor-based coding and facial similarity perception.
\newblock {\em Perception 26}, 1\_suppl (1997), 250.

\bibitem{lyons2007dimensional}
{\sc Lyons, M.~J.}
\newblock Dimensional affect and expression in natural and mediated
  interaction.
\newblock {\em arXiv preprint arXiv:1707.09599\/} (2007).

\bibitem{lyons2000linked}
{\sc Lyons, M.~J., Morikawa, K., and Akamatsu, S.}
\newblock A linked aggregate code for processing faces.
\newblock {\em Pragmatics \& Cognition 8}, 1 (2000), 63--81.

\bibitem{martine1866martine}
{\sc Martine, A.}
\newblock {\em Martine's Hand-book of Etiquette: And Guide to True Politeness}.
\newblock Dick \& Fitzgerald, 1866.

\bibitem{phillips2000feret}
{\sc Phillips, P.~J., Moon, H., Rizvi, S.~A., and Rauss, P.~J.}
\newblock The feret evaluation methodology for face-recognition algorithms.
\newblock {\em IEEE Transactions on pattern analysis and machine intelligence
  22}, 10 (2000), 1090--1104.

\bibitem{prabhu2020large}
{\sc Prabhu, V.~U., and Birhane, A.}
\newblock Large image datasets: A pyrrhic win for computer vision?
\newblock {\em arXiv preprint arXiv:2006.16923\/} (2020).

\bibitem{russell1980circumplex}
{\sc Russell, J.~A.}
\newblock A circumplex model of affect.
\newblock {\em Journal of personality and social psychology 39}, 6 (1980),
  1161.

\bibitem{russell2003core}
{\sc Russell, J.~A.}
\newblock Core affect and the psychological construction of emotion.
\newblock {\em Psychological review 110}, 1 (2003), 145.

\bibitem{schlosberg1952description}
{\sc Schlosberg, H.}
\newblock The description of facial expressions in terms of two dimensions.
\newblock {\em Journal of experimental psychology 44}, 4 (1952), 229.

\bibitem{solon2019facial}
{\sc Solon, O.}
\newblock Facial recognition’s ‘dirty little secret’: Millions of online
  photos scraped without consent.
\newblock {\em NBC News\/} (2019).
\newblock [Online; accessed 27-August-2020].

\bibitem{srinivasan2018cross}
{\sc Srinivasan, R., and Martinez, A.~M.}
\newblock Cross-cultural and cultural-specific production and perception of
  facial expressions of emotion in the wild.
\newblock {\em IEEE Transactions on Affective Computing\/} (2018).

\bibitem{tedone2019from}
{\sc Tedone, G.}
\newblock From spectacle to extraction. and all over again.
\newblock {\em unthinking photography\/} (2020).
\newblock [Online; accessed 27-August-2020].

\bibitem{tomkins1962affect}
{\sc Tomkins, S.~S.}
\newblock {\em Affect imagery consciousness: Volume I: The positive affects},
  vol.~1.
\newblock Springer publishing company, 1962.

\bibitem{wang2018deep}
{\sc Wang, Y., and Kosinski, M.}
\newblock Deep neural networks are more accurate than humans at detecting
  sexual orientation from facial images.
\newblock {\em Journal of personality and social psychology 114}, 2 (2018),
  246.

\bibitem{wundt1897outline}
{\sc Wundt, W.}
\newblock {\em Outlines of psychology.}
\newblock Leipzig: Wilhelm Engelmann, 1897.

\bibitem{zhang1998comparison}
{\sc Zhang, Z., Lyons, M., Schuster, M., and Akamatsu, S.}
\newblock Comparison between geometry-based and gabor-wavelets-based facial
  expression recognition using multi-layer perceptron.
\newblock In {\em Proceedings Third IEEE International Conference on Automatic
  face and gesture recognition\/} (1998), IEEE, pp.~454--459.

\bibitem{zuboff2019age}
{\sc Zuboff, S.}
\newblock {\em The Age of Surveillance Capitalism: The Fight for a Human Future
  at the New Frontier of Power: Barack Obama's Books of 2019}.
\newblock Profile Books, 2019.

\end{thebibliography}

\end{document}